\documentclass[conference]{IEEEtran}

\usepackage{graphicx}
\graphicspath{{/figures_TMM/}}
\usepackage{algorithm}
\usepackage{algorithmic}
\usepackage{amsfonts,amsmath,amssymb,array,balance,caption,cite,color,dsfont,epstopdf,float,longtable,multirow,mathtools,subcaption,subeqnarray,xcolor,xkeyval}
\newcommand{\st}{\mathop{\rm s.t.}\limits}

\makeatletter

\newcommand{\Rmnum}[1]{\expandafter\@slowromancap\romannumeral #1@}
\newcommand{\tabincell}[2]{\begin{tabular}{@{}#1@{}}#2\end{tabular}}  
\makeatother

\begin{document}
	\title{Dynamic Video Streaming in Caching-enabled Wireless Mobile Networks}
	\author
	{\IEEEauthorblockN{C. Liang and S. Hu}
	\IEEEauthorblockA{Depart. of Systems and Computer Eng., Carleton Univ., Ottawa, ON, Canada}
	
	}

\maketitle
	\begin{abstract}
Recent advances in software-defined mobile networks (SDMNs), in-network caching, and mobile edge computing (MEC) can have great effects on video services in next generation mobile networks. In this paper, we jointly consider SDMNs, in-network caching, and MEC to enhance the video service in next generation mobile networks. With the objective of maximizing the mean measurement of video quality, an optimization problem is formulated. Due to the coupling of video data rate, computing resource, and traffic engineering (bandwidth provisioning and paths selection), the problem becomes intractable in practice. Thus, we utilize dual-decomposition method to decouple those three sets of variables. Extensive simulations are conducted with different system configurations to show the effectiveness of the proposed scheme.
	\end{abstract}
	
	\IEEEpeerreviewmaketitle
	
\section{Introduction}\label{sec:intro}
The video service is replacing voice and other applications to become the fundamental service in mobile networks \cite{WZR14,HMK16,Schoolar2015Mobile}. Moreover, high-definition (HD) videos (e.g., 720p, 1080p, and beyond) that request at least 5-20 Mbps user data rate will be ubiquitous \cite{Huawei2016Opening}, which will bring great challenges to the design and operation of next generation mobile networks (e.g., 5G and beyond). To address these challenges, recent advances of information and communications technologies can be explored, such as \emph{software-defined mobile networks} (SDMNs) \cite{chen2015software}, \emph{in-network caching} \cite{LYZ15} and \emph{mobile edge computing} (MEC) \cite{hu2015mobile}.\par 
SDMNs have been proposed to fully support SDN design in wireless networks, which enable the programmability in mobile networks so that the complexity and the cost of networks can be reduced \cite{chen2015software}.  With the programmability, SDN is considered as a promising candidate to enhance traffic engineering \cite{mendiola2016survey}. The success of the utilizing SDMNs for traffic engineering depends critically on our ability to jointly provision the backhaul and radio access networks (RANs) for the traffic \cite{farmanbar2014traffic,liao2014min,dao2015handling}.\par
Another promising technology, in-network caching, as one of the key features of \emph{information-centric networking} (ICN), can efficiently reduce the duplicate content transmission in networks\cite{ADIKO01}. Particularly, caching content (e.g., videos) at mobile edge node (e.g., base stations (BSs) and routers) has been proposed as one of the key enablers in next generation mobile networks \cite{paschos2016wireless,liu2016caching,he2016big,HLY17}.\par 
MEC has attracted great interest recently as computational resources are moved closer to users, which can efficiently improve the quality of service (QoS) for applications that require intensive computations (e.g., video processing and tracking) \cite{hu2015mobile}. With the widely employed HTTP adaptive streaming, such as Dynamic Adaptive Streaming over HTTP (DASH) (e.g., Google and Netflix), the video client can request the proper quality level adaptively according to the network throughput. MEC that deploys computing servers at BSs of the radio access network (RAN) can proactively optimize the delivery of videos by transcoding videos to desired qualities according to network conditions \cite{pedersen2015enhancing,tran2016collaborative}.\par 
Although some works have been done on SDMNs, in-network caching and MEC separately, jointly considering these new technologies to enhance the video service has been largely ignored in the existing research. In this paper, we jointly consider Mobile Edge Computing and Caching (MECC)-enabled SDMNs to enhance the video service in next generation mobile networks. Specifically, in the proposed framework, a SDN controller is deployed to steer the bandwidth provisioning and traffic paths selection while assisting nodes to perform edge computing and video quality selection. In addition, popular videos can be stored at caches of network nodes (e.g.,
router and BSs). We design an efficient mechanism that jointly considers  network-assisted video rate adaptation, bandwidth provisioning in traffic engineering, and computing resource scheduling in MEC.\par
The rest of this paper is organized as follows. Section \ref{sec:system} introduces the system model and formulates the presented problem. Section \ref{sec:proposed} describes the proposed algorithms and the corresponding analysis. Simulation results are discussed in Section \ref{sec:sim}. Finally, we conclude this study in  Section \ref{sec:conclusions}.
\section{System Model and Problem Formulation}\label{sec:system}
	In this section, we	present the system model of video streaming, mobile network, MEC and caching. The problem is formulated after the related assumptions are given.\par
	\subsection{Network Model}\label{sec:network}
		We assume that users watch streaming videos selected from a source library $\mathcal{F}$ (e.g., Youtube or Netflix) and each video streaming is served by one data flow. For simplicity, one user can only watch one video at the same time, which leads us to use the notation $i$ to index the flow and the corresponding user. Each video file $f\in\mathcal{F}$ is encoded at a finite number of different quality levels (resolutions) $q\in\{1,...,Q\}$, which is similar to DASH service. We assume that each level $q$ requires a minimum data rate $v_q$ (bps) to support smooth playback. Practically, $v_q$ depends on video coding schemes and the video content, which are varying with time. Nevertheless, since the purpose of our paper is to maximize the mobile network performance dynamically, we can consider the required data rate is a fixed value $v_q$ when we do the scheduling.For simplicity, vertical handover \cite{MYL04,YK07,MYL07} is not considered in this paper.	To evaluate the gain of the video quality, the measure of each video quality $q$ is defined as $s_q$.\par
	%
		In this paper, we consider a MECC-enabled heterogeneous network (HetNet) with the backhaul network and the radio access network (RAN). The backhaul network is assumed to be a mesh network connecting SBSs, MBSs and Gateways (GWs) by wired links with fixed capacities. Users are connected to BSs with wireless channels sharing total $W$ (Hz) radio spectrum resource. This considered network is modeled by a directed graph $G(\mathcal{N},\mathcal{L})$. $\mathcal{N}$ includes network nodes (GWs and BSs) formed a set $\mathcal{J}$ and users formed a set $\mathcal{I}$. $\mathcal{L}$ comprised with wired and wireless links denoted by sets $\mathcal{L}^{wd}$ and $\mathcal{L}^{wl}$, respectively. $m_l\in\mathcal{N}$ and $n_l\in\mathcal{N}$ are used to denote the destination node and the source node of  link $l$, respectively.\par
		If link $l$ is a wired link, it is assumed to provide a fixed bandwidth capacity $B_l$. If link $l$ is a wireless link, the capacity depends on the ratio of the radio resource that the network allocates to this link. In this paper, to simply our analysis, we do not consider any advanced interference management and power allocation schemes. Thus, by using the Shannon bound, the spectrum efficiency of  wireless link $l$ is defined as $\gamma_l = \log\left(1+g_{n_lm_l}p_{n_l}/\sigma_0\right)$, where $g_{n_lm_l}$ is the large-scale channel gain that includes pathloss and shadowing between the transmission node $n_l$ (the source of  link $l$) and the receiving node $m_l$ (the destination of  link $l$). We deploy the same model used in \cite{ye2013user} to calculate the pathloss and apply shadowing.  $p_{n_l}$ (Watt/Hz) is the normalized transmission power on  link $l$. The fixed equal power allocation mechanism is used, which means transmission power $p_{n_l}$ is the same for all frequencies. $\sigma_0$ is the power spectrum density of additive white Gaussian noise. Accordingly, the achievable data rate capacity of  link $l$ is $R_l = W\gamma_l$.\par
		Each data flow can be spitted to multiple paths as the user is assumed to be served by multiple BSs through  BS cooperation \cite{liao2014min} or multistream carrier aggregation. Moreover,  demanded videos can be potentially retrieved from any nodes (GWs, BSs or the source server) where matched data are found, which means each user can download the data of video from different places. For example, the data flow of user 1 is split to two paths where one is from content source server to MBS 1 then to users 1 and another is directly from SBS 2.\par
		The network is equipped with caching and computing functions on network nodes. We assume a subset $\mathcal{F}_j$ of $\mathcal{F}$ is stored at node $j$. It should be noted that  node $j$ always caches the highest quality $Q$ of  video file $f$ so that it can be transcoding to a lower quality. As we mentioned above, if video file $f_i$ demanded by  user $i$ is found at node $j$, namely $f_i\in\mathcal{F}_j$,  node $j$ becomes a candidate source. To indicate a hitting event between user $i$ and  node $j$, we define $h_{ij}=1$ and $v_{ij}=v_i$ if $f_i\in\mathcal{F}_j$. $v_{ij}=v_i$ means the video requested by  user $i$ can be potentially fully provided by  node $j$ and $h_{ij}=1$ means a successful hitting event. If $f_i\notin\mathcal{F}_j$, $h_{ij}=+\infty$ that means there is no hitting event (infinite resource is required to process this video) and $v_{ij}=0$ means  node $j$ cannot response the video request $i$.\par 
		If $h_{ij}=1$ and  node $j$ is selected as one of source nodes of flow $i$, the video data needs to be transcoded to the required quality level except that the highest quality level is selected. However, unlike the powerful computing resource at the source server (e.g., the data center), due to the computing resource at each node, limited tasks can be activated at the same time. Similar to \cite{pedersen2015enhancing}, we define the maximum mobile computing capacity as the number of encoded bits that can be processed per second, denoted by $C_j$ (bps). For example, a 500 Mbps computing capacity means 20 concurrent video processing tasks are allowed.
	\subsection{Problem Formulation}\label{sec:formulation}
	The purpose of the considered problem is to find an optimal video quality level for each user with considering network resources and the cached video distribution. We define a binary variable $x_{qi}\in\{0,1\}$ as the resolution indicator of user $i$. Specifically, if the $q$-th resolution of the video is selected by  user $i$, $x_{qi}=1$; otherwise, $x_{qi}=0$.\par
	To support video services demanded by users, an optimal path set for all data flows should be found by solving the proposed algorithm. Denote $\mathcal{P}_{i}$ as a path set including all candidate paths for  user $i$ and $\mathcal{P}_{ij}$ as a subset of $\mathcal{P}_{i}$ including all candidate paths starting from  node $j$. A path $p^{k}_{ij}\in\mathcal{P}_{i}$ means the $k$-th path of  flow $i$ starting from  node $j$ and the corresponding data rate of this path is denoted by $r^{k}_{ij}$ if $p^{k}_{ij}$ is selected. Thus the achievable data rate of  flow $i$ is $\sum_{p^{k}_{ij}\in\mathcal{P}_{i}}r^{k}_{ij}$ that is the aggregated rate of all selected paths.\par
	As we mentioned in above, the computing resource on each node needs to be scheduled to video transcoding tasks. Thus, we define a binary variable $y_{ij}\in\{0,1\}$ as the computing task assignment indicator. If $y_{ij}=1$, node $j$ is able to trans-code the video demanded by user $i$ to a desired quality level; otherwise, the video data cannot be retrieved from  node $j$.
	To improve the whole network utility by maximizing the overall mean gain of videos, the SDN controller performs traffic engineering to assist users adaptively selecting optimal video quality levels. Thus, the proposed problem can be formed as follows.
		\begin{subequations}\label{eq:proOrg}
			\begin{equation}\label{eq:objOrg}
			\max_{\mathbf{X},\mathbf{R},\mathbf{Y}} \quad U(\mathbf{X})=\frac{1}{|\mathcal{I}|}\sum_{i\in\mathcal{I}} \sum_{q=1}^{Q} s_qx_{qi}
			\end{equation}
			subject to 
			\begin{equation}\label{eq:ctrRegion}
			\begin{split}
			r^k_{ij}\in\Re^+, \forall i,j,k,\\
			x_{qi},y_{ij}\in \{0,1\},\forall q, i, j,k,
			\end{split}
			\end{equation}
		\begin{equation}\label{eq:ctrQuality}
			\sum_{q=1}^{Q} x_{qi}=1, \forall i\in \mathcal{I},
		\end{equation}
		\begin{equation}\label{eq:ctrResourceC}
		\sum_{i\in\mathcal{I}}y_{ij}h_{ij}c_i \leq C_{j}, \forall j\in \mathcal{J},
		\end{equation}
			\begin{equation}\label{eq:ctrResourceWd}
				\sum_{p^k_{ij}\in\mathcal{P}_l} r^k_{ij} \leq B_l, \forall l \in \mathcal{L}^{wd},
			\end{equation}
			\begin{equation}\label{eq:ctrResourceWl}
				\sum_{l \in \mathcal{L}^{wl}} \frac{\sum_{p^k_{ij}\in \mathcal{P}_l}{r^k_{ij}}}{\gamma_{l}} \leq W,
			\end{equation}
			\begin{equation}\label{eq:ctrService}
				\sum_{p^k_{ij}\in\mathcal{P}_i} r^k_{ij} = \sum_{q}v_{qi}x_{qi}, \forall i\in \mathcal{I},
			\end{equation}
			\begin{equation}\label{eq:ctrContent}
				\sum_{p^k_{ij}\in\mathcal{P}_{ij}}r^k_{ij} \leq  v_{ij}\left(x_{Qi} + y_{ij}\right), \forall i\in \mathcal{I},j\in \mathcal{J},k\in \mathcal{K},
			\end{equation}	
		\end{subequations}
where $\{x_{qi}\}$, $\{y_{ij}\}$ and $\{r^k_{ij}\}$ are elements of $\bf{X}$, $\bf{Y}$ and $\bf{R}$, respectively. 	Constraint (\ref{eq:ctrQuality}) reflects that only one resolution level can be selected for one user. The computing capacity on each node $j$ is specified by the constraint (\ref{eq:ctrResourceC}) where $c_i$ (bps) is the computing resource required for transcoding  video $f_i$. The flow conservation law (FCL) of traffic engineering is claimed by constraints (\ref{eq:ctrResourceWd}) and (\ref{eq:ctrResourceWl}) where $\mathcal{P}_l$ is the set of paths that pass  link $l$. (\ref{eq:ctrResourceWd}) means the allocated data rate of  link $l$ for all passing path should be less than the link capacity. As the radio resource is shared by the whole RAN, (\ref{eq:ctrResourceWl}) enforces that the total allocated spectrum cannot exceed the available spectrum bandwidth. The demand constraint for every video flow is given by the (\ref{eq:ctrService}). The constraint (\ref{eq:ctrContent}) requires that any candidate path starting from  node $j$ can be selected only when it has the content in the cache ($h_{ij}=1$) and the computing resource is assigned to transcode this content ($y_{ij}=1$) or the highest quality is selected ($x_{Qi}=1$).\par
	Unfortunately, problem (\ref{eq:proOrg})  is difficult to solve and implement. Firstly, the mix integer variables result in the problem a mix-integer linear problem (MILP) that generally is NP-complete. Moreover, video resolution, path selection, and resource scheduling are decided by different layers, network nodes and perform in different time scales. Lastly, the necessary exchange of local information about the network and links affects the performance as overheads are introduced.
\section{Proposed Scheme}\label{sec:proposed}
	In this section, dual-decomposition method is deployed to simplify  problem (\ref{eq:proOrg}).\par
	This paper aims to give an efficient scheme to help video clients to select appropriate video resolutions while conducting resource scheduling to provision bandwidth and process video data. The network needs to transfer some information to assist users and nodes when they perform video selection and processing so that the optimal network utility can obtained. \par
	We firstly define independent local feasible sets $\varPi_{{r}}$, $\varPi_x$, and $\varPi_y$ for variables ${\bf{R}}$, $\bf{X}$ and $\bf{Y}$, respectively. Those feasible regions only subject to constraints that include one type of variables, which are shown as
		 \begin{equation}\label{def:setFeaR}
			\varPi_{r} = \left\{ \{{r}^k_{ij}\} \left| \begin{array}{l}
			\Re^+,(\ref{eq:ctrResourceWd}),(\ref{eq:ctrResourceWl})
			\end{array} \right.\right\}.
		\end{equation}
		\begin{equation}\label{def:setFeaX1}
			\varPi_x = \left\{ \{x_{qi}\} \left| \begin{array}{l}
			\{0,1\},(\ref{eq:ctrQuality})
			\end{array} \right.\right\}.
		\end{equation}
		\begin{equation}\label{def:setFeaX2}
			\varPi_y = \left\{ \{y_{ij}\} \left| \begin{array}{l}
			\{0,1\},(\ref{eq:ctrResourceC})
			\end{array} \right.\right\}.
		\end{equation} 
Fortunately, the coupled constraints are (\ref{eq:ctrService}) and (\ref{eq:ctrContent}). Thus, by relaxing constraints (\ref{eq:ctrService}) and (\ref{eq:ctrContent}) with dual variables $\{\mu_i\}$ and $\{\lambda_{ij}\}$\footnote{Dual variables can be interpreted as costs of bandwidth and computing}, the Lagrangian can be shown as:
		\begin{equation}\label{eq:defDualLarg}
		\begin{split}
			\max_{\mathbf{X},\mathbf{R},\mathbf{Y}}&\quad	U(\mathbf{X})
			+\sum_{i\in\mathcal{I}}\mu_{i}\left[\sum_{p^k_{ij}\in\mathcal{P}_{i}}r^k_{ij}- \sum_{q=1}^{Q}v_{qi}x_{qi}\right]\\
			&-\sum_{i\in\mathcal{I},j\in\mathcal{J}}\lambda_{ij}\left[\sum_{p^k_{ij}\in\mathcal{P}_{ij}}r^k_{ij}-v_{ij}\left(x_{Qi}+y_{ij}\right)\right]\\
			\st&\quad \mathbf{X}\in\varPi_x,\mathbf{Y}\in\varPi_y,\mathbf{R}\in\varPi_{r}.
		\end{split}
		\end{equation}	
		Thus, the original problem has been separated to two levels of optimization that are higher level for updating dual variables and low level for finding dual functions \cite{palomar2006tutorial}. Accordingly, the dual problem (DP) then is:
			\begin{equation}\label{eq:proDual}
			\textbf{DP}:\min_{\mu\in\mathbb{R},\lambda\in\mathbb{R}^+}\quad D(\mu,\lambda)=g_x(\mu,\lambda)+g_r(\mu,\lambda)+g_y(\lambda)
			\end{equation} 
where $g_x(\mu,\lambda)$, $g_r(\mu,\lambda)$, and $g_y(\lambda)$ are dual functions obtained as the maximum value of the Lagrangians solved in following problems (\ref{eq:proRes}), (\ref{eq:proRate}) and (\ref{eq:proCom}) for given $\{\mu_i\}$ and $\{\lambda_{ij}\}$.
			\begin{equation}\label{eq:proRes}
				g_x(\mu,\lambda)=\sup_{x_{qi}\in\varPi_x}\left\{\begin{array}{l}
				U(\mathbf{X})-\sum_{i\in\mathcal{I}}^{I}\mu_{i}\sum_{q}v_{qi}x_{qi}\\
				+\sum_{i\in\mathcal{I},j\in\mathcal{J}}\lambda_{ij}v_{ij}x_{Qi}
				\end{array}\right\},
			\end{equation}
			\begin{equation}\label{eq:proRate}
				g_r(\mu,\lambda)=\sup_{r^k_{ij}\in\varPi_r}\left\lbrace \sum_{i\in\mathcal{I}}\sum_{p^k_{ij}\in\mathcal{P}_{i}}\left(\mu_{i}-\lambda_{ij}\right)r^k_{ij}\right\rbrace,
			\end{equation}
			\begin{equation}\label{eq:proCom}
			g_y(\mu,\lambda)=\sup_{y_{ij}\in\varPi_y}\left\lbrace \sum_{i\in\mathcal{I},j\in\mathcal{J}}\lambda_{ij}v_{ij}y_{ij}\right\rbrace,
			\end{equation}		
		It is observed that $D(\mu,\lambda)$ is not a differentiable function due to the binary variables and candidate path sets. Thus, we can deploy subgradient method to solve the dual problem (\ref{eq:proDual}). Obviously, a sub-gradient of  problem (\ref{eq:proDual}) for  $\lambda_{ij}$ is:
			$z^{\lambda}_{ij}=\sum_{p^k_{ij}\in\mathcal{P}_{ij}}r^k_{ij}-v_{ij}\left(x_{Qi}+y_{ij}\right)$,
		and for $\mu_{i}$ is
			$z^{\mu}_{i}=\sum_{p^k_{ij}\in\mathcal{P}_{i}}r^k_{ij}- \sum_{q}v_{qi}x_{qi}$.
		According to dual decomposition, we thus can update $\mu_{i}$ and $\lambda_{ij}$ based on:
		\begin{equation}\label{eq:updateMu}
			\mu_{i}^{[t+1]}=\mu_i^{[t]}-\tau^{[t]}_{\mu}z^{\mu}_{i},
		\end{equation}
		and
		\begin{equation}\label{eq:updateLambda}
		\lambda_{ij}^{[t+1]}=\left[\lambda_{ij}^{[t]}-\tau^{[t]}_{\lambda}z^{\lambda}_{ij}\right]^+,
		\end{equation}
		where $\tau ^{[t]}_{\mu}$ and $\tau^{[t]}_{\lambda}$ are the length of step at  iteration step $[t]$.\par
		Thus, if we are able to solve the inner problems (\ref{eq:proRes}), (\ref{eq:proRate}) and (\ref{eq:proCom}) in each iteration, the SDN controller can update dual variables and transfer them to nodes and users to assist them to find optimal solutions of their own variables $x_{qi}$, $r^k_{ij}$, and $y_{ij}$. In the remaining of this section, algorithms will be given to solve problems (\ref{eq:proRes}), (\ref{eq:proRate}) and (\ref{eq:proCom}).
%
	%
	Observe that problem (\ref{eq:proRes}) can be decoupled to users where the local problem of each user is shown as
		\begin{equation}\label{eq:proResLocal}
			\begin{split}
				\max_{x_{qi}\in\{0,1\}}&\quad	\sum_{q=1}^{Q}s_qx_{qi}-\mu_{i}\sum_{q}v_{qi}x_{qi}
				+\sum_{j\in\mathcal{J}}\lambda_{ij}v_{ij}x_{Qi}\\
				\st&\quad \sum_{q=1}^{Q} x_{qi}=1.
			\end{split}
		\end{equation}
	The above problem can be solved with effortless due to that only one quality level can be selected. Thus, each user only needs to select the level maximizing the utility.\par 
	Similar to  problem (\ref{eq:proRes}),  problem (\ref{eq:proCom}) also can be decoupled to each node $j$ as follows.
		\begin{equation}\label{eq:proComLocal}
			\begin{split}
			\max_{y_{ij}\in\{0,1\}}\quad \sum_{i\in\mathcal{I}}\lambda_{ij}v_{ij}y_{ij}\\
			\st\quad \sum_{i\in\mathcal{I}}y_{ij}h_{ij}v_i \leq C_{j},
			\end{split}
		\end{equation}
	Obviously, this is a 0-1 knapsack problem that is usually NP-complete. Thus, BnB method or dynamic programming can be used to solve this problem, but they are both computationally intensive and might not be practical for large-scale problem. Therefore, firstly, to ease the size of this problem so that common methods can be used, we form a set $\mathcal{I}^+_j$ including every user who has non-zero gain $\lambda_{ij}v_{ij}$ or finite $h_{ij}$. Formally, $\mathcal{I}^+_j:=\{i|\lambda_{ij}v_{ij}>0, h_{ij}<\infty\}$. It is easy to see that $y_{ij}=0$ if $i\notin \mathcal{I}^+_j$. Thus, we only need to consider users that in $\mathcal{I}^+_j$, which leads to a reduction of the problem size.\par
	This problem (\ref{eq:proRate}) is easy to solve theoretically as it is a linear problem. However, wired backhaul and radio access links are involved in this problem, which leads the solution hard to achieve in practice. Denote the data rate of path $p^k_{ij}$ decided by the wired backhaul network is $\check{r}^{k}_{ij}$ and its peer decided by the RAN is $\hat{r}^{k}_{ij}$. Furthermore, the wired backhaul network can be decoupled to links, as capacities of wired links usually are independent with each other. $\check{r}^{k,l}_{ij}$ denotes that the data rate of path $p^k_{ij}$ allocated by  link $l$. It should be noted that $\check{r}^{k,l}_{ij}$ does not mean $p^k_{ij}$ passing  link $l$. $\check{r}^{k,l}_{ij}$ can be considered the opinion for $p^k_{ij}$ given by  link $l$. Actually, all local varialbes of $r^{k}_{ij}$ is the opinion or recommendation from one part of the whole network. By defining $\check{r}^{k,l}_{ij}$ and $\hat{r}^{k}_{ij}$,  problem (\ref{eq:proRate}) can be revised as
		\begin{equation}\label{eq:proRateDC}
		\begin{split}
			\max_{\check{r}^{k,l}_{ij},\hat{r}^{k}_{ij}\in\mathbb{R}^+}&\quad\frac{1}{2L^{wd}}\sum_{l\in{L}^{wd}}\sum_{i\in\mathcal{I}}\sum_{p^k_{ij}\in\mathcal{P}_{i}}\left(\mu_{i}-\lambda_{ij}\right)\check{r}^{k,l}_{ij}\\
			&+\frac{1}{2}\sum_{i\in\mathcal{I}}\sum_{p^k_{ij}\in\mathcal{P}_{i}}\left(\mu_{i}-\lambda_{ij}\right)\hat{r}^{k}_{ij}\\
			\st&\quad\sum_{i\in\mathcal{I}}\sum_{p^k_{ij}\in\mathcal{P}_{i}} \alpha^{k,l}_{ij}\check{r}^{k,l}_{ij} \leq B_l, \forall l \in \mathcal{L}^{wd},\\
			&\quad\sum_{l\in\mathcal{L}^{wl}} \frac{\sum_{i\in\mathcal{I}}\sum_{p^k_{ij}\in\mathcal{P}_{i}}{\alpha^{k,l}_{ij}\hat{r}^{k}_{ij}}}{\gamma_{l}} \leq W,\\
			&\check{r}^{l,k}_{ij} = \hat{r}^{k}_{ij},\forall i,j,k,l.
		\end{split}
	\end{equation}
	where $L^{wd}=|\mathcal{L}^{wd}|$ is the number of wired backhaul links. By ussing dual decomposition method, the equality constraints in \ref{eq:proRateDC} can be decoupled to each links or RAN. Thus, problem (\ref{eq:proRateDC}) can be sovled locally at each wired link or RAN similar to the method we use to decouple video adaptation and computing. 
	\section{Simulation Results and Discussions}
	\label{sec:sim}
	In the simulation, we consider a cellular network, consisting of one MBS, 15 SBSs and 15 active users, that covers a 250m-by-250m area. Transmission with single antenna for both transmitter and receiver is considered in our paper. The remaining simulation parameters are summarized in Table \ref{tbl:parameters}. We assume that the total 1000 videos are in the video library $\mathcal{F}$. In this paper, to adapt ubiquitous HD videos in next generation mobile networks, we refer the measure to U-vMOS proposed in \cite{Schoolar2015Mobile,Huawei2016Opening}. Each video $f$ can be encoded to 6 levels with constant bit rate (CBR), and has the same length of 600 seconds.\par
	Files in $\mathcal{F}$ have been sorted according to the popularity. We assume that the popularity of each video being requested follows a Zipf distribution with exponent 0.56. The $f$-th most popular video has a request probability of $(f^{-0.56})/(\sum_{f'=1}^{|\mathcal{F}|}f'^{-0.56})$. The default cache capacity $S_n$ of a MBS is 200 video files, and default cache capacity of a SBS is 100 video files with the highest resolution, which leads the hitting rate of around 50\% at MBS  and 40\% at SBSs. Least Frequently Used (LFU) caching policy is used at the MEC server to place/replace videos in caches, which means each BS stores the most $S_n$ popular video files.\par
	The MEC computing capability of a MBS is set to 150 Mbps that is equivalent to processing 6 videos simultaneously. Due to limitations of SBSs, the computing performance of a MEC server at SBS is only set to 50 Mbps equivalent to 2 videos.\par
		\begin{table}[tp]
			\centering
			\caption{Network parameters settings}
			\label{tbl:parameters}
			\begin{tabular}[width=0.4\textwidth]{l|l}
				\hline
				Network parameters              & value \\ 
				\hline\hline
				frequency bandwidth (MHz)       & 20  \\
				transmission power profile		& \tabincell{l}{SISO with maximum power;\\49dBm (MBS), 20dBm (SBS)} \\
				propagation profile				& \tabincell{l}{pathloss:L(distance)=34+40log(distance);\\lognormal shadowing: 8dB;\\no fast fading}\\
				power density of the noise		& -174 dBm/Hz\\
				backhaul capacity (Mbps)        & \tabincell{l}{MBS to GW:$100$;\\ SBS to MBS: $50$}\\
				\hline	
			\end{tabular}
		\end{table}
	\begin{figure}[t]
		\centering
		\begin{subfigure}{0.24\textwidth}
			\includegraphics[width=\textwidth]{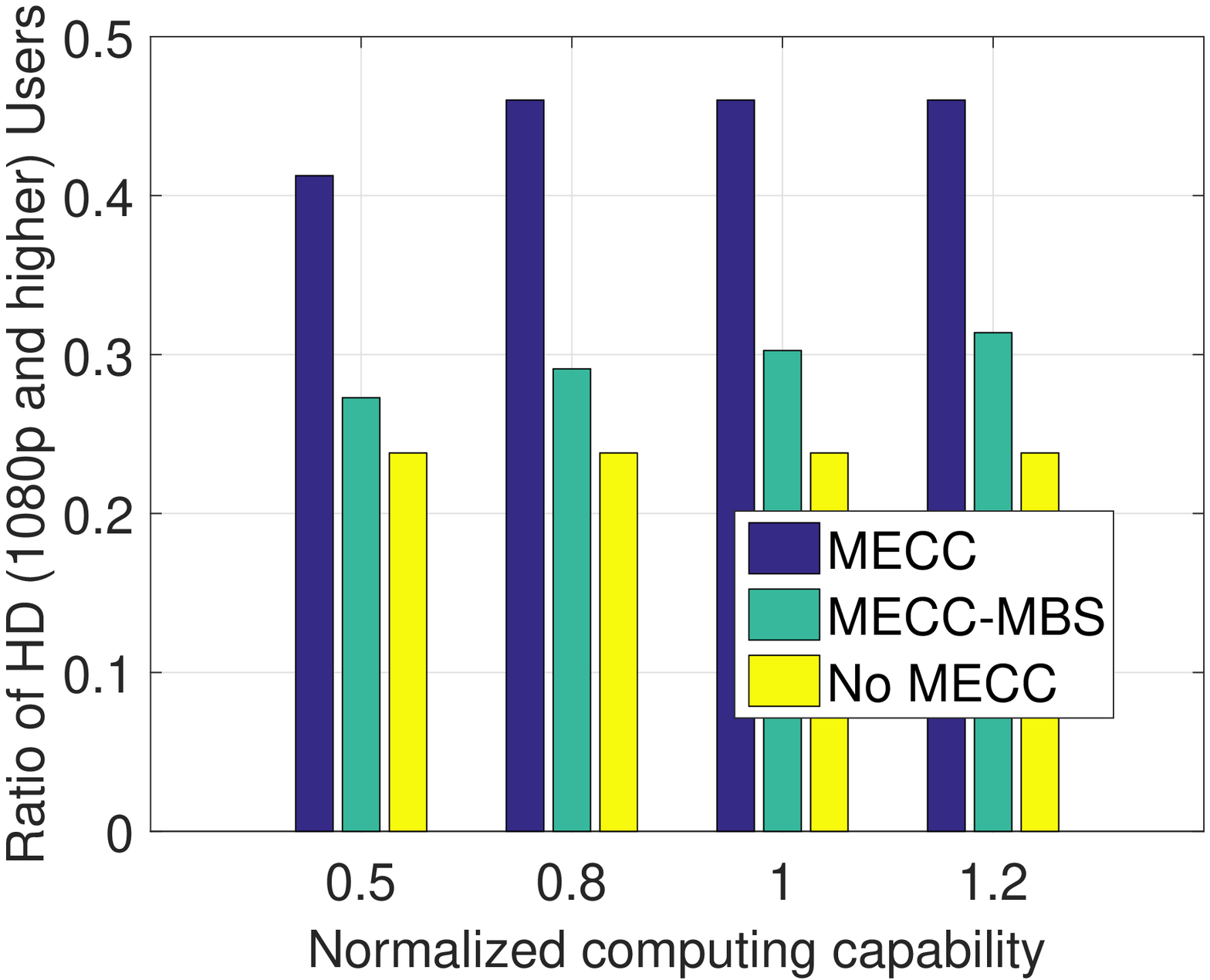}
			\caption{Computing resource}
			\label{fig:HDmec}
		\end{subfigure}
		\begin{subfigure}{0.24\textwidth}
			\includegraphics[width=\textwidth]{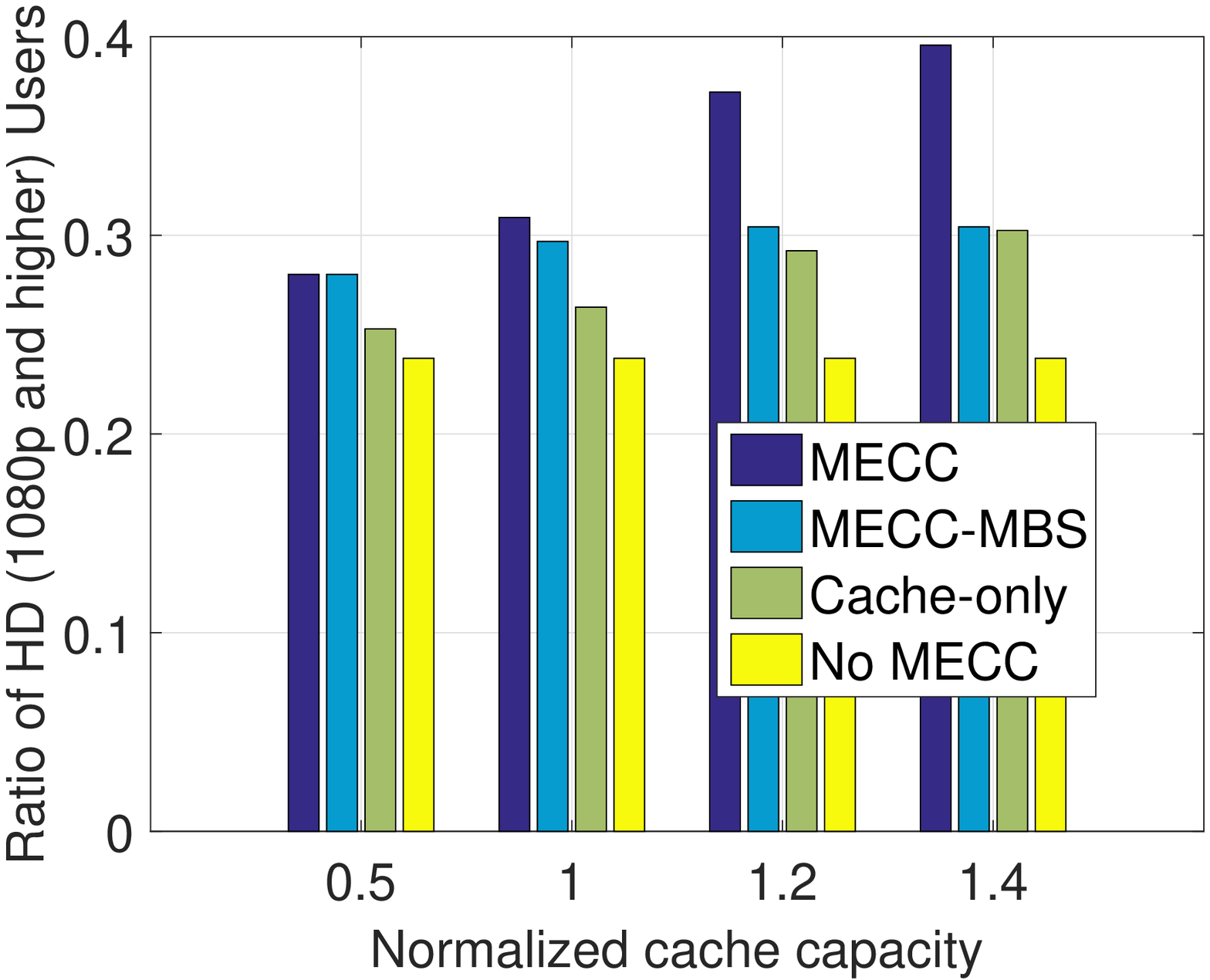}
			\caption{Cache storage capacity}
			\label{fig:HDcache}
		\end{subfigure}
		\caption{The distribution of video resolutions with different network setups.}\label{fig:HD}
	\end{figure}
	It is also necessary to discuss the distribution of video resolutions of our proposed scheme. As shown in Fig. \ref{fig:HD}, overall, it can be seen that the ratio of 1080p and higher resolutions decreases with the load of the networks, and increases with the available network resource. The proposed scheme gets much higher ratio, up to 15\%, on the 1080p and higher resolutions compared to cache-only case and no MECC case. Some observations can be made from Fig. \ref{fig:HD}. Firstly, the MECC improves the performance of the network-assist video rate adaptation significantly on HD videos. Secondly, if only MBSs can provide MEC, the gain of the proposed scheme is not as good as a traditional HetNet. Moreover, compared to traditional networks, in-network caching also can enhance the quality of video services. Furthermore, the capability of MEC servers may not always be the bottleneck of the system.

\section{Conclusions}\label{sec:conclusions}
	In this paper, we jointly studied the  video rate adaptation problem in a MEC-enable SDMN where in-network caching was deployed. an optimization problem was formulated with the objective of maximizing the mean video measurement of a HetNet. Dual-decomposition method has been utilized to decouple video data rate, computing resource, and traffic engineering (bandwidth provisioning and path selection) so that those variables could be obtained independently. Simulation results were presented to show that  our proposed scheme can significantly improve the mean video measurement.

	\ifCLASSOPTIONcaptionsoff
	\newpage
	\fi

\section*{Acknowledgment}
This work was supported in part by the Natural Sciences and Engineering Research Council of Canada and Huawei Technologies Canada CO., LTD.

	\balance
	\bibliographystyle{IEEEtran}

	\bibliography{Liang_References}

\end{document}